\documentclass[twocolumn,english,aps,pra,superscriptaddress,showpacs,tightenlines]{revtex4-1}
\usepackage{amsmath}
\usepackage{graphicx}
\usepackage{amssymb}
\usepackage{CJK}

\begin{document}

\begin{CJK*}{GBK}{song}

\title{Spontaneous emission of a moving atom in a waveguide of rectangular cross section}

\author{Jing \surname{Zeng} }
\affiliation{Key Laboratory of Low-Dimensional Quantum Structures and Quantum Control of
Ministry of Education, Department of Physics and Synergetic Innovation
Center of Quantum Effects and Applications, Hunan Normal University,
Changsha 410081, China}
\author{Jing \surname{Lu} }
\affiliation{Key Laboratory of Low-Dimensional Quantum Structures and Quantum Control of
Ministry of Education, Department of Physics and Synergetic Innovation
Center of Quantum Effects and Applications, Hunan Normal University,
Changsha 410081, China}
\author{Lan \surname{Zhou} }
\thanks{Corresponding author}
\email{zhoulan@hunnu.edu.cn}
\affiliation{Key Laboratory of Low-Dimensional Quantum Structures and Quantum Control of
Ministry of Education, Department of Physics and Synergetic Innovation
Center of Quantum Effects and Applications, Hunan Normal University,
Changsha 410081, China}

\begin{abstract}
We study the spontaneous emission (SE) of an excited two-level nonrelativistic system (TLS) interacting with
the vacuum in a waveguide of rectangular cross section. All TLS's transitions and the center-of-mass motion
of the TLS are taken into account. The SE rate and the carried frequency of the emitted photon for the TLS
initial being at rest is obtained, it is found in the first order of the center of mass (c.m.) that the frequency
of the emitted photon could be smaller or larger than the transition frequency of the TLS but the SE
rate is smaller than the SE rate $\Gamma_{f}$ of the TLS fixed in the same waveguide. The SE rate and the
carried frequency of the emitted photon for the TLS initial being moving is also obtained in the first order
of the c.m.. The SE rate is larger than $\Gamma_{f}$ but it is independent of the initial momentum. The carried
frequency of the emitted photon is creased when it travels along the direction of the initial momentum and
is decreased when it travels in the opposite direction of the initial momentum.
\end{abstract}

\maketitle

\end{CJK*}\narrowtext

\section{Introduction}

Photons are desirable for distributing information and transferring
entanglement in quantum networks. With the demanding to build a device in a
quantum network for controlling single photons, single quantum emitters
(Hereafter, we will often use the word ``atom'' instead of ``quantum
emitter.'') would be desirable since there is no direct interaction among
photons. An two-level system (TLS) act as a quantum switch in a one
dimension (1D) waveguide with linear dispersion relation~\cite{Fans} or a 1D
coupled-resonator waveguide (CRW)~\cite{ZLPRL08}; A V-type or ladder-type
atom functions as the frequency converter for single photons~\cite%
{ZLfconvert} in a 1D waveguide; A cyclic or a $\Lambda$-type atom works as a
multichannel quantum router~\cite{ZLQrouter}. Atoms are widely proposed to
act as quantum nodes in extended communication networks and scalable
computational devices~\cite%
{zlZeno,Zheng,ZLMZI,GongPRA78,LawPRA78,SPT13,ZLQrouter2,PRL116,attenuator,MTcheng,Longo,Alexanian,TShiSun}%
. Photons propagating along the network, are confined in a 1D waveguide.
With the mode volume decreased, the coupling strength of atoms to the
waveguide is enhanced. The atomic spontaneous emission (SE) has played an
important role on controlling photon transport in quantum network.

SE is a result of the electromagnetic interaction between an atom and a
quantum field. It is a process of the following: a system initially in an
excited state relaxed to its lower state and emitted a quanta of energy to
its surrounding vacuum field, which carries away the difference in energy
between the two levels. Different from atoms in a cavity or a free space,
atoms used to control single photons in quantum networks interact with its
surrounding electromagnetic (EM) field which is confined in a 1D waveguide.
SE has been widely studied in free space, a semi-infinite or infinite 1D
waveguide, however, most works focus on 1D waveguide without a cross
section, and the atom under study is usually assumed to be stationary. A
stationary atom possesses an undetermined kinetic energy according to the
Heisenberg's uncertainty relation, and thus, the stationary-atom model is
too simple. In this paper, we study the spontaneous radiative decay of an
atom moving in an infinite waveguide of rectangular cross section. We
analyze the interaction of an initially excited TLS with the waveguide in
vacuum. When an initially excited TLS spontaneously emits a photon into the
waveguide, the atomic center-of-mass (c.m.) experiences a change in momentum
space. The dynamical behavior of the TLS is studied by taking the
quantization of TLS's momentum and position into account.

This paper is organized as follows. In Sec.~\ref{Sec:2}, we introduce the
model and establishe the notation. In Sec.~\ref{Sec:3}, we derive the
relevant equations describing the dynamics of the system for the case of the
TLS being initially excited and the waveguide mode in the vacuum state. In
Sec.~\ref{Sec:4}, the spontaneous emission rate has been presented for TLS
initially at rest and moving respectively. We make a conclusion in Sec.~\ref%
{Sec:5}.


\section{\label{Sec:2}model setup}

\begin{figure}[tbp]
\includegraphics[clip=true,height=4cm,width=7cm]{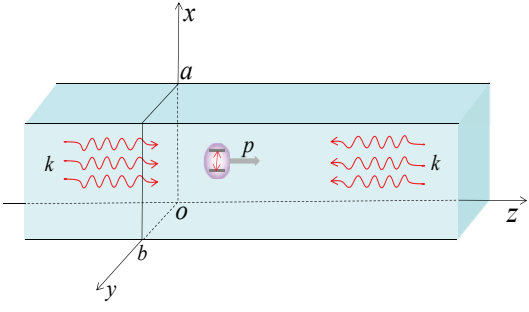}
\caption{(Color online) Schematic illustration for the two-level atom inside
a infinite waveguide of rectangular cross section.}
\label{fig:1}
\end{figure}

The system we studied is shown in Fig.~\ref{fig:1}. A waveguide made of
ideal perfect conducting walls is formed from surfaces at $x=0$, $x=a$, $y=0$%
, $y=b$, and is placed along the $z$ axis.
There are two types of modes for the field in the waveguide: TE modes whose
electric field has no longitudinal component, and TM modes whose magnetic
field has no longitudinal component. Let $\vec{k}=\left(
k_{x},k_{y},k\right) $ be the wave vector. The relations $k_{x}=m\pi /a$ and
$k_{y}=n\pi /a$ with positive integers $n,m\ $can be imposed by the
condition that the tangential components of the electric field vanish at all
the conducting wall, however, there is no constraint on $k$. Therefore, the
waveguide allows a continuous range of frequencies described by the
dispersion relation
\begin{equation}
\omega _{mnk}=\sqrt{c^{2}k^{2}+\Omega _{mn}^{2}},  \label{A-1}
\end{equation}%
where $c$ is the speed of light in vacuum, the cutoff frequency for a
traveling wave $\Omega _{mn}=\pi c\sqrt{m^{2}/a^{2}+n^{2}/b^{2}}$. We note
that $m$ and $n$ cannot both be zero. If $a>b$, TE$_{10}$ is the lowest
guiding mode for the waveguide \cite{EMtextb}, and the lowest TM modes occur
for $m=1$, $n=1$. Obviously, the waveguide modes form a one-dimensional
continuum. Each guiding mode provides a quantum channel for photons to
travel from one location to the other.

An TLS with mass $M$ and the transition frequency $\omega _{A}$ is inside
the waveguide. It has two internal states: the upper level $\left\vert
e\right\rangle $ and the lower level $\left\vert g\right\rangle $, which
introduce the the rising (lowering) atomic operator $\hat{\sigma} _{+}\equiv
\left\vert e\right\rangle \left\langle g\right\vert$ ($\hat{\sigma}
_{-}\equiv \left\vert g\right\rangle \left\langle e\right\vert $). We assume
that the atom moves only along the $z$ axis, which means the atom has
unchanged position along the $xy$ plane. The atomic momentum and position
are characterized by the operators $\hat{p}_z$ and $\hat{\vec{r}}=\left(
x_{0},y_{0},\hat{z}\right)$ respectively, where $x_{0}$ and $y_{0}$ are
constant. The momentum and position of the atomic c.m. obey canonical
commutation relations. The free Hamiltonian for the TLS is described by
\begin{equation}
H_{s}=\frac{\hat{p}_{z}^{2}}{2M}+\omega _{A}\hat{\sigma} _{+}\hat{\sigma}
_{-}.  \label{A-2}
\end{equation}
We consider the atomic electric dipole is oriented along the $z$ direction,
which means that the TLS only interacts with the TM$_{mn}$ modes. Since the
number $\left( m,n,k\right) $ specifies the mode function of this air-filled
metal pipe waveguide, we label the annihilation operator for each TM guiding
mode by $a_{mnk}$. The free Hamiltonian for the waveguide is described by
\begin{equation}
H_{f}=\sum_{mn}\int_{-\infty }^{\infty }dk\text{ }\omega _{mnk}\hat{a}%
_{mnk}^{\dag }\hat{a}_{mnk}  \label{A-3}
\end{equation}%
The interaction between the TLS and field via the dipole coupling in the
rotating-wave approximation reads
\begin{equation}
H_{I}=\sum_{mn}\int_{-\infty }^{\infty }dkg_{mnk}(\hat{\sigma} _{-}\hat{a}%
_{mnk}^{\dag }e^{-ik\hat{z}}-\hat{\sigma} _{+}\hat{a}_{mnk}e^{ik\hat{z}})
\label{A-4}
\end{equation}%
where the coupling strength%
\begin{equation}
g_{mnk}=\frac{id\Omega_{mn}}{\sqrt{\pi A\epsilon_{0}\omega_{mnk}}} \sin\frac{%
m\pi x_0}{a}\sin\frac{n\pi y_0}{b}  \label{A-5}
\end{equation}%
Here, $\epsilon _{0}$ the permittivity of free space, $d$ the magnitude of
the transition dipole moment of the TLS and assumed to be real, $A=ab$ the
area of the rectangular cross section.

The TLS located at $x_{0}=a/2$ and $y_{0}=b/2$ decouples to the TM$_{mn}$
guiding mode with even integer $m$ or $n$. The total system are described by
Hamiltonian $H=H_{s}+H_{f}+H_{I}$.


\section{\label{Sec:3} dynamic of the moving atom}

Let us introduce the operators of the excited number and the atom-field
momentum
\begin{subequations}
\label{B-1}
\begin{eqnarray}
N_{e}&=& \hat{\sigma}_{+}\hat{\sigma}_{-}+\sum_{mn}\int_{-\infty}^{\infty}dk%
\text{ }\hat{a}_{mnk}^{\dag }\hat{a}_{mnk}, \\
N_{p} &=& \hat{p}_{z}+\sum_{mn}\int_{-\infty}^{\infty}dk\text{ }\hbar k\hat{a%
}_{mnk}^{\dag}\hat{a}_{mnk},
\end{eqnarray}
respectively. It can be found that both $N_{e}$ and $N_{p}$ are commuted
with the total Hamiltonian of the system, thus, the excited number and
atom-field momentum are conservative quantities. It is well-known that an
stationary TLS being excited initially will evolves into a superposition of
itself and the states in which the TLS is unexcited and has released a
photon into the field, however, when the quantization of the atomic momentum
and position are taking into account, the conservation of momentum indicates
that the c.m. motion of the atom is changed by the recoil caused by the
emission of photons.

The initial wave function of the system we consider is in a product state $%
|\psi(0)\rangle=|e,p_z,0\rangle$, where $|p_z\rangle$ is an eigenstate of
the momentum operator $\hat{p}_z$ corresponding to the eigenvalue $p_z$ and
the field is in the vacuum state. The time evolution of the state $%
|\psi_t\rangle$ is described by the Schr\"{o}dinger equation. Via the
Fourier transformation, the Green operator $\hat{G}=(q-\hat{H})^{-1}$
relates the state $|\psi_t\rangle$ to the initial state $|\psi_0\rangle$,
which yields
\end{subequations}
\begin{equation}
|\psi_t\rangle=-\frac{1}{2\pi i}\int_{-\infty}^{+\infty}dq\frac{e^{-iqt}}{q-%
\hat{H}}\left\vert \psi_0\right\rangle.  \label{B-2}
\end{equation}
The probability for finding the atom in its excited state equals to the
probability for finding the TLS in its initial state $|\psi_0\rangle$, which
is denoted by $P_{A} =\left\vert \left\langle \psi_0 \right.
\left\vert\psi_t \right\rangle \right\vert ^{2}$. Since the atom-field
coupling energy $H_I$ is small compared with the free energy $%
H_0=H_{s}+H_{f} $, the Green operator $\hat{G}$ can be expanded into a
series of ascending powers of $H_I$. By defining the free Green operator $%
\hat{G}_0=(q-\hat{H}_0)^{-1}$, the probability $P_A$ is rewritten as
\begin{equation}
P_{A}\left( t\right) =\left\vert \left\langle \psi_0\right| \hat{G}%
_0\sum_{l=0}^{+\infty}\left(H_I\hat{G}_0\right)^l\left\vert\psi_0
\right\rangle \right\vert ^{2}.  \label{B-3}
\end{equation}
Since the photon's emission transform the external state of the TLS from $%
|p_z\rangle$ to $|p_z-\hbar k\rangle$, we observe that
\begin{subequations}
\label{B-4}
\begin{eqnarray}
\left\langle \psi_0\right| \left(H_I\hat{G}_0\right)^{2l+1}\left\vert\psi_0
\right\rangle &=& 0, \\
\left\langle \psi_0\right| \left(H_I\hat{G}_0\right)^{2l}\left\vert\psi_0
\right\rangle &=& \left(\left\langle \psi_0\right| H_I\hat{G}_0H_I\hat{G}%
_0\left\vert\psi_0 \right\rangle\right)^l.
\end{eqnarray}
Then the amplitude for finding the TLS in its initial state reads
\end{subequations}
\begin{equation}
A\left( t\right) =-\frac{1}{2\pi i} \int_{-\infty}^{+\infty}dq\frac{e^{-iqt}%
}{q-p_{z}^{2}\diagup(2M)-\omega_{A}-B},  \label{B-5}
\end{equation}
where the quantity $B$ is defined as%
\begin{equation}
B=\sum_{mn}\int_{-\infty}^{\infty}dk\frac{\left\vert g_{mnk}\right\vert ^{2}%
}{q-\left( p_{z}-\hbar k\right) ^{2}\diagup(2M)-\omega_{mnk}}  \label{B-6}
\end{equation}
Eq.(\ref{B-5}) can be solved by the residue theorem, but the solution of the
equation
\begin{equation}
0=q-p_{z}^{2}\diagup(2M)-\omega_{A}-B  \label{B-7}
\end{equation}
has to be found. Since Eq.(\ref{B-7}) is a transcendental equation involving
integral, the approximations are used to give an analytic discussion. In
this following discussion, for the sake of simplification, we denote the
transversally confined propagating modes which couple to the atom as TM$_{j}$
with $j=\left( m,n\right) $ according to the ascending order of the cutoff
frequencies.


\section{\label{Sec:4} Spontaneous emission}

The roots of the denominator in Eq. (\ref{B-7}) can be splitted into a sum
of the singular and principal value parts. The principal part is merely
included into redefinition of the energy. The singular part gives the decay
rate due to the coupling to field. If the coupling strength $g_{mnk}$ is
small, the solution of Eq.(\ref{B-7}) can be expanded into a series of
ascending powers of $g_{jk}$. Under the second-order approximation about the
weak coupling, the atomic decay is dominantly exponential with rate
\begin{equation}
\Gamma =2\pi\sum_{j}\int dk\left\vert g_{jk}\right\vert ^{2}\delta\left(
\omega_{A}+\frac{p_z^{2}}{2M}-\omega_{jk}-\frac{(p_z-k)^{2}}{2M}\right)
\label{c-0}
\end{equation}
The delta function expresses the conservation of energy before and after the
emission. We express the integral in Eq.~(\ref{c-0}) in terms of the
frequency
\begin{equation}
\Gamma =\sum_{j}\int_{\Omega _{j}}^{\infty } \frac{d\omega4\pi \omega
\left\vert g_{j\omega }\right\vert ^{2}}{c\sqrt{\omega ^{2}-\Omega _{j}^{2}}}%
\delta\left(\omega_{A}-\omega_{jk}+\frac{p_zk}{M}-\frac{k^{2}}{2M}\right)
\label{c-1}
\end{equation}
by using the dispersion relation, where the coupling strength
\begin{equation}
g_{j\omega }=\frac{id\Omega _{mn}}{\sqrt{\pi A\epsilon _{0}\omega }}\sin
\frac{m\pi x}{a}\sin \frac{n\pi y}{b}.  \label{c-3}
\end{equation}

\subsection{the TLS being initially at rest}

We now assume that the TLS is initially at rest, i.e., $p_{z}=0$. The
spontaneous decay rate becomes
\begin{equation}
\Gamma _{R}=\sum_{j}\int_{\Omega _{j}}^{\infty }d\omega \frac{4\pi \omega
\left\vert g_{j\omega }\right\vert ^{2}}{c\sqrt{\omega ^{2}-\Omega _{j}^{2}}}%
\delta \left( \omega _{A}-\omega -\frac{\omega ^{2}-\Omega _{j}^{2}}{2Mc^{2}}%
\right) .  \label{c-2}
\end{equation}%
From the delta function, the frequency that the emitted photon carried is
obtained
\begin{equation}
\omega _{R}=\sqrt{\left( Mc^{2}\right) ^{2}+2Mc^{2}\omega _{A}+\Omega
_{j}^{2}}-Mc^{2}\text{,}  \label{c-4}
\end{equation}%
After some algebra, we obtain the spontaneous rate for a TLS initially at
rest
\begin{eqnarray}
\Gamma _{R} &=&\sum_{mn}\frac{2d^{2}\Omega _{mn}^{2}\sin ^{2}\frac{m\pi x}{a}%
\sin ^{2}\frac{n\pi y}{b}}{Ac\epsilon _{0}\sqrt{\omega _{A}^{2}-\Omega
_{mn}^{2}}}  \label{c-5} \\
&&\times \sqrt{\frac{2Mc^{2}\omega _{A}+2\left( Mc^{2}\right) ^{2}}{\sqrt{%
\left( Mc^{2}\right) ^{2}+2Mc^{2}\omega _{A}+\Omega _{mn}^{2}}}+2Mc^{2}}
\notag
\end{eqnarray}%
It can be found that the modal profile affects on the decay rate via
location of the TLS. If the atom is located at $x_{0}=a/2$ and $y_{0}=b/2$,
no photons are radiated into the TM$_{mn}$ guiding mode with even integer $m$
or $n$ since the guiding mode are standing waves in the transverse
direction. As the initial atomic energy $\omega _{A}$ approaches one of the
cutoff frequencies, the TLS loses its energy very quickly. The more
transverse modes interact with the TLS, the faster the TLS decays. The mass
also contributes to the spontaneous emission rate due to the recoil of the
TLS.

To comparing with the TLS fixed inside a waveguide with a rectangular cross
section, we consider the mass is larger and keep the first order of $M^{-1}$%
. Then the spontaneous rate reduces to%
\begin{equation}
\Gamma _{R}=\left( 1-\frac{3\omega _{A}}{4Mc^{2}}\right) \sum_{mn}\Gamma
_{mn}^S  \label{c-6}
\end{equation}%
which is smaller than spontaneous rate of the TLS fixed in the vacuum of the
waveguide with a rectangular cross section. Here,
\begin{equation}
\Gamma _{mn}^S=\frac{4d^{2}\Omega _{mn}^{2}}{Ac\epsilon _{0}}\frac{\sin ^{2}%
\frac{m\pi x}{a}\sin ^{2}\frac{n\pi y}{b}}{\sqrt{\omega _{A}^{2}-\Omega
_{mn}^{2}}}  \label{c-7}
\end{equation}%
is the decay rate of a stationary TLS due to its interacting with the vacuum
of the TM$_{mn}$ transverse mode. The frequency of the emitted photon
approximately reads
\begin{equation}
\omega _{R}=\omega _{A}+\frac{\Omega _{mn}^{2}-\omega _{A}^{2}}{2Mc^{2}}.
\label{c-8}
\end{equation}%
It shows that the cutoff frequencies around the transition frequency $\omega
_{A}$ determines whether $\omega _{w}^{R}$ is smaller or larger than the
transition frequency of the TLS. If the mass of the TLS is so large that the
recoil can be neglect, we recover the decay rate of the stationary TLS and
the frequency $\omega _{A}$ of the emitted photon.

\subsection{the TLS initially moving}

When the atom is not initially at rest, i.e., $p_{z}\neq 0$, the roots of
the following equations
\begin{equation}
0=\omega_{A}-\omega\pm\frac{p_{z}}{Mc}\sqrt{\omega^{2}-\Omega_{j}^{2}}-\frac{%
\omega^{2}-\Omega_{j}^{2}}{2Mc^{2}}  \label{c-9}
\end{equation}
determine the frequency of the emitted photon as well as the value of
spontaneous rate. We note that Hamiltonian (\ref{A-2}) is written in the
nonrelativistic region where the motion of center of mass of the TLS is very
slower than light, so we apply the perturbation approach to Eq.(\ref{c-9})
to find the analytic solutions,
\begin{equation}
\omega =\omega_{0}+\lambda\omega_{1}+\cdots,  \label{c-10}
\end{equation}
Here, $\lambda$ is a continuously varying parameter ranging from zero to
unity. $\omega_{0}$ is of the zeroth order in $(Mc)^{-1}$, $\omega_{1}$ is
of the first order in $(Mc)^{-1}$. We now substitute Eq. (\ref{c-10}) into
Eq. (\ref{c-9}) and retain only terms up to the first order in $(Mc)^{-1}$.
We approximately obtained frequencies of the emitted photon
\begin{subequations}
\label{c-11}
\begin{eqnarray}
\omega_{+} & =\omega_{A}+\frac{p_{z}}{Mc}\left( \omega_{A}^{2}-\Omega
_{j}^{2}\right) -\frac{\omega_{A}^{2}-\Omega_{j}^{2}}{2Mc^{2}} \\
\omega_{-} & =\omega_{A}-\frac{p_{z}}{Mc}\left( \omega_{A}^{2}-\Omega
_{j}^{2}\right) -\frac{\omega_{A}^{2}-\Omega_{j}^{2}}{2Mc^{2}}
\end{eqnarray}
The nonvanishing initial C.M. motion of the TLS splits the frequency of the
emitted photon in Eq.(\ref{c-8}) into two. For the emitted photon traveling
along the same direction of the initial TLS momentum, its carried frequency
is increased. For the emitted photon traveling in the opposite direction of
the initial TLS momentum, its carried frequency is decreased. The emission
of the right-going and left-going photon along the $z$ axis leads to a loss
of TLS's energy with rate
\end{subequations}
\begin{equation}
\Gamma_M=2\pi\sum_{j}\left( \frac{\omega_{+}\left\vert
g_{j\omega_{+}}\right\vert ^{2}}{c\sqrt{\omega^{2}_{+}-\Omega_j^{2}}}+\frac{%
\omega_{-}\left\vert g_{j\omega_{-}}\right\vert ^{2}}{c\sqrt{%
\omega^{2}_{-}-\Omega_{j}^{2}}}\right).  \label{c-12}
\end{equation}
For a moving atom, the mass, momentum and the cutoff frequencies all
contribute to the decay rate. As $\omega^{\pm}_w$ approaches one of the
cutoff frequencies, the TLS loses its energy very quickly. The more
transverse modes interact with the TLS, the faster the TLS decays. For a
much larger mass, the decay rate reads
\begin{equation}
\Gamma_M=\sum_{mn}\Gamma_{mn}^{S}\left( 1+\frac{\omega_{A}}{2Mc^{2}}\right)
\end{equation}
in the first order of $M^{-1}$ , where the contribution of the momentum
disappeared. However, the spontaneous emission rate definitely increased an
amount.


\section{\label{Sec:5}conclusion}

We consider a TLS of c.m. interacting with a waveguide of rectangular cross
section. The spontaneous emission of an atom in vacuum is studied. Both atom
and field variables are fully quantized, so that the atomic recoil on the
emission of radiation are automatically included in the calculation. Under
the second order approximation of the weak coupling, the SE rate and the
frequency of the emitted photon for the atom initial being at rest or moving
are obtained. It is shown that the modal profile affects on the decay rate
via location of the TLS, the more transverse modes interact with the TLS,
the faster the TLS decays. For the TLS initial being at rest, the transition
frequency $\omega_A$ of the TLS determines the speed of the SE rate. As $%
\omega_A$ approach any of the cutoff frequencies, the TLS decays fast. In
the first order of the c.m., the SE rate is smaller than the SE rate of the
TLS fixed in the vacuum of the same waveguide, however, the frequency of the
emitted photon could be larger or smaller than the initial energy $\omega_A$
of the total system. For the TLS initial being moving, the carried frequency
of the emitted photon is creased when it travels along the direction of the
initial momentum of the TLS and is decreased when it travels in the opposite
direction of the initial momentum of the TLS. The SE rate is larger than the
SE rate of the TLS fixed in the vacuum of the same waveguide but it is
independent of the initial momentum in the first order of the c.m..

\begin{acknowledgments}
This work was supported by NSFC Grants No. 11374095, No. 11422540, No.
11434011, No. 11575058.
\end{acknowledgments}

\end{document}